\documentclass[manuscript]{aastex}
\usepackage{emulateapj5}

\usepackage{epsfig}
\usepackage{amsmath}
\usepackage{amssymb}
\usepackage{natbib}
\usepackage{graphicx}

\shorttitle{Spin equilibrium of Aql X--1}
\shortauthors{Bhattacharyya}

\begin{document}

\title{Application of a new method to study the spin equilibrium of Aql X--1: the possibility of gravitational radiation}

\author{Sudip Bhattacharyya}
\affil{Department of Astronomy and Astrophysics, Tata Institute of
  Fundamental Research, 1 Homi Bhabha Road, Colaba, Mumbai 400005,
  India; sudip@tifr.res.in}

\begin{abstract}
Accretion via disks can make neutron stars in low-mass X-ray binaries (LMXBs) fast spinning, and some
of these stars are detected as millisecond pulsars.
Here we report a practical way to find out if a neutron star in a transient LMXB has reached the spin equilibrium
by disk--magnetosphere interaction alone, and if not, to estimate this spin equilibrium frequency. 
These can be done using specific measurable source luminosities, such as the
luminosity corresponding to the transition between the accretion and propeller phases,
and the known stellar spin rate.
Such a finding can be useful to test if the spin distribution of millisecond pulsars,
as well as an observed upper cutoff of their spin rates,
can be explained using disk--magnetosphere interaction alone, or additional spin-down
mechanisms, such as gravitational radiation, are required. Applying our method,
we find that the neutron star in the transient LMXB Aql X--1 has not yet reached 
the spin equilibrium by disk--magnetosphere interaction alone.
We also perform numerical computations, with and without gravitational radiation, to study the spin evolution 
of Aql X--1 through a series of outbursts and to constrain its properties.
While we find that the gravitational wave emission from Aql X--1 cannot be established with certainty, our numerical results
show that the gravitational radiation from Aql X--1 is possible, with a $1.6\times10^{37}$ g cm$^2$
upper limit of the neutron star misaligned mass quadrupole moment.
\end{abstract}

\keywords{accretion, accretion disks --- methods: analytical --- methods: numerical ---
pulsars: general --- pulsars: individual (Aql X--1) --- X-rays: binaries}

\section{Introduction}\label{Introduction}

Millisecond pulsars (MSPs) are fast-spinning neutron stars. They are believed to attain 
a high spin frequency ($\nu$), typically of several hundred Hz, by accretion-induced angular momentum transfer 
in their low-mass X-ray binary (LMXB) phase \citep{RadhakrishnanSrinivasan1982, Alparetal1982,
WijnandsKlis1998, ChakrabartyMorgan1998, Archibaldetal2009, Papittoetal2013, deMartinoetal2013, 
Bassaetal2014}.
However, the details of such angular momentum transfer from a companion star via
an accretion disk is not fully understood yet. For example, it is not clear why the spin 
distribution of MSPs is what it is \citep{Watts2012, PatrunoWatts2012}, and 
why these spin frequencies cut off above $\nu_{\rm cut} \sim 730$ Hz 
\citep{Chakrabartyetal2003, Chakrabarty2005, Patruno2010, FerrarioWickramasinghe2007,
Hessels2008, Papittoetal2014}, which is well below the neutron star breakup spin rates
\citep{Cooketal1994, Bhattacharyyaetal2016}.

The most basic reason to explain the cutoff frequency involves the equilibrium 
spin frequency $\nu_{\rm eq}$, which we briefly describe here.
For a spinning, magnetic neutron star, which accretes from a geometrically thin Keplerian
accretion disk, the disk inner edge should be located where the rate of angular momentum 
removal from the disk by the stellar magnetic field begins to exceed the viscous stress.
This gives the disk inner edge radius or the magnetospheric radius \citep[e.g., ][]{Wang1996}
\begin{equation}\label{rm}
	r_{\rm m} = \xi \left(\frac{\mu^4}{2 G M \dot M^2}\right)^{1/7}.
\end{equation}
Here, $M$ is the stellar gravitational mass (hereafter, mass), $\dot M$ 
is the accretion rate, $\mu~(= BR^3)$ is the stellar 
magnetic dipole moment, $B$ is the stellar surface dipole magnetic field, $R$ is the 
stellar radius and $\xi$ is an order of unity constant. 
Note that $\xi$ depends on the disk--magnetosphere interaction and the magnetic pitch 
at the inner edge of the disk \citep{Wang1996}. Another important radius,
the corotation radius, where the disk Keplerian angular velocity is the same as the
stellar angular velocity, is given by
\begin{equation}
	r_{\rm co} = \left(\frac{GM}{4\pi^2 \nu^2}\right)^{1/3}.
\end{equation}
When $r_{\rm co} > r_{\rm m}$, i.e., in the accretion phase, 
accretion in an LMXB happens, the neutron star spins up by a net positive torque,
and $r_{\rm co}$ decreases. On the other hand, when $r_{\rm co} < r_{\rm m}$, 
i.e., in the so-called ``propeller regime," accreted matter
is partially driven away from the system \citep{Illarionov1975,Ustyugova2006,
DAngeloetal2010,BhattacharyyaChakrabarty2017} at the expense of the stellar angular momentum, resulting 
in a stellar spin-down by a net negative torque and 
an increase of $r_{\rm co}$. Therefore, $r_{\rm co}$ tends to become $r_{\rm m}$
by a self-regulated mechanism. This spin equilibrium corresponds to a frequency
(obtained by $r_{\rm m} = r_{\rm co}$ condition):
\begin{equation}\label{equilibrium}
\nu_{\rm eq} = \frac{1}{2\pi}\sqrt\frac{GM}{r_{\rm m}^3} = 
\frac{1}{2^{11/14}\pi\xi^{3/2}}\left(\frac{G^5 M^5 \dot
	M^3}{\mu^6}\right)^{1/7},
\end{equation}
which is the maximum spin frequency a neutron star can attain by disk--magnetosphere 
interaction, and which could explain the above-mentioned observed upper cutoff frequency.
Moreover, many accreting neutron stars may have already attained the $\nu_{\rm eq}$ 
value through the disk--magnetosphere interaction, and if so, the
spin equilibrium frequency may largely explain the  spin distribution of MSPs
\citep[e.g., ][]{LambYu2005, Patrunoetal2012b}.

However, it is not certain and is a topic of current debate 
whether spin equilibrium due to disk--magnetosphere
interaction alone can explain the spin distribution and cutoff. For example,
for a high accretion rate and/or a low $B$ value, $\nu_{\rm eq}$ could be much higher than
$\nu_{\rm cut}$. Therefore, a spin-down due to gravitational radiation
may be required to explain the cutoff spin frequency and the  spin distribution
\citep{Bildsten1998, Anderssonetal1999, Chakrabartyetal2003}.
On the other hand, most of the X-ray MSPs accrete mass with a low average rate, and
hence there is a possibility that their spin rates could be explained with the 
disk--magnetosphere interaction alone, as $\nu_{\rm eq}$ ($\propto \dot{M}^{3/7}$; see 
Equation~\ref{equilibrium}) could be low \citep[see, for example, ][]{LambYu2005}. 
But almost all the X-ray MSPs are transient sources \citep{Watts2012, PatrunoWatts2012},
and recently \citet{BhattacharyyaChakrabarty2017} have shown that a neutron
star in a transient can be spun up to a frequency several times higher than that of
a neutron star in a persistent source for the same long-term average $\dot{M}$.
This is because, while Equation~\ref{equilibrium} works for persistent sources,
a different expression of the spin equilibrium frequency should be used for transient sources.
Therefore, \citet{BhattacharyyaChakrabarty2017} 
have suggested that an additional spin-down, for example due to gravitational radiation,
may be required to explain the observed spin distribution and $\nu_{\rm cut}$.

In order to test the above-mentioned possibilities for transients, it is very important to devise a way
(1) to find out if a neutron star in an LMXB has reached the spin equilibrium 
by disk--magnetosphere interaction alone, and if not, 
then (2) to estimate the spin equilibrium frequency.
It will be particularly useful if this method works even with a small number of 
known parameter values.
In this paper, we provide such a method for transient LMXBs to achieve both the 
above-mentioned goals. We also apply this 
method for a prolifically outbursting transient neutron star LMXB Aql X--1
\citep[e.g., ][]{Gungoretal2014}. We also perform
numerical computation of the evolution of Aql X--1 through a series of outbursts, and 
report an upper limit of the stellar misaligned mass quadrupole moment, which causes the
gravitational radiation.

\section{Torques and spin equilibrium for transients}\label{previous}

\citet{BhattacharyyaChakrabarty2017} have reported a crucial effect of transient 
accretion on the spin-up of MSPs, which was not recognized earlier.
The new method presented in our paper uses this effect, and hence, in this section,
we briefly discuss the torques on and spin equilibrium of
neutron stars in transient systems from \citet{BhattacharyyaChakrabarty2017}.

In our numerical calculations, we use the following expressions of torques on a spinning 
neutron star due to disk--magnetosphere interaction \citep{Rappaportetal2004,BhattacharyyaChakrabarty2017}:
\begin{equation}\label{Torque7}
        N = \dot M \sqrt{GMr_{\rm m}} + \frac{\mu^2}{9 r_{\rm m}^3}\left[2\left(\frac{r_{\rm m}}{r_{\rm co}}\right)^3-6\left(\frac{r_{\rm m}}{r_{\rm co}}\right)^{3/2}+3\right]
\end{equation}
for the accretion phase, and
\begin{equation}\label{Torque8}
        N = -\eta \dot M \sqrt{GMr_{\rm m}} -\frac{\mu^2}{9 r_{\rm m}^3}\left[3-2\left(\frac{r_{\rm co}}{r_{\rm m}}\right)^{3/2}\right]
\end{equation}
for the propeller phase.
In each expression, the first term is the material contribution, while the second term is due to 
the disk--magnetosphere interaction (see \citet{BhattacharyyaChakrabarty2017} for derivation).
In the propeller phase, for a relatively small 
range of $\dot{M}$ values, $r_{\rm m}$ is close to $r_{\rm co}$.
For this range of $\dot{M}$, only an unknown fraction of the accreting matter could be expelled from the 
system, and the rest of the matter could accumulate and eventually fall on the neutron star
in a cyclic manner \citep[e.g., ][]{DAngeloetal2010}. 
This could introduce some uncertainty in the numerical results, when Equation~\ref{Torque8}
is used. But note that this could happen only for a small fraction of the propeller phase 
duration for transients, as $\dot{M}$ evolves considerably \citep{BhattacharyyaChakrabarty2017}.
Moreover, we have also checked that such a cyclic accretion for a limited time 
can have a small effect (at most a few percent) on the 
long-term spin evolution of neutron stars. Besides, the uncertainty in the
material torque due to an unknown fraction of matter ejected in the propeller phase
is accounted for by an order of unity positive constant $\eta$ \citep[Equation~\ref{Torque8};
see also ][]{BhattacharyyaChakrabarty2017}. Therefore, it is reasonable to
use Equation~\ref{Torque8}.

For our analytical calculations, for the sake of simplicity, we use a torque expression
\begin{equation}\label{Torque9} 
        N = \frac{{\rm d}J}{{\rm d}t} = \pm A \dot{M}^{6/7},
\end{equation}
which can be approximated from Equations~\ref{Torque7} and \ref{Torque8}, 
as shown in \citet{BhattacharyyaChakrabarty2017}. This approximation
gives at most a few percent error in the spin evolution.
Here $J$ and $A$ are the stellar angular momentum and a positive constant, respectively, and the
positive sign corresponds to the accretion phase, while the negative sign corresponds to the 
propeller phase. Note that $A$ is a function of $M$, $\xi$, $B$, $R$ and a constant $\beta$, 
where $0 \le \beta \le 1$ \citep[see Equation~18 of ][]{BhattacharyyaChakrabarty2017}.

In order to compare the theoretical results with observations, it is convenient 
to use the source luminosity $L$ instead of $\dot{M}$. It is reasonable to assume
$L \propto \dot{M}$ \citep[e.g., ][]{Patrunoetal2012b}, and hence we can rewrite
Equation~\ref{Torque9} as
\begin{equation}\label{Torque10} 
       N = \frac{{\rm d}J}{{\rm d}t} = \pm \alpha L^{6/7},
\end{equation}
where $\alpha$ is a function of $A$ and the proportionality constant between $L$ and $\dot{M}$.
However, note that the  $L - \dot{M}$ relation is expected to be somewhat different in accretion and in propeller phases
(as the outburst decay profile should steepen when a source enters in the propeller phase),
and hence the $L$-value of one of these phases should be suitably scaled.

\begin{figure}[h]
\hspace{-1.0cm}
\includegraphics*[width=18.0cm]{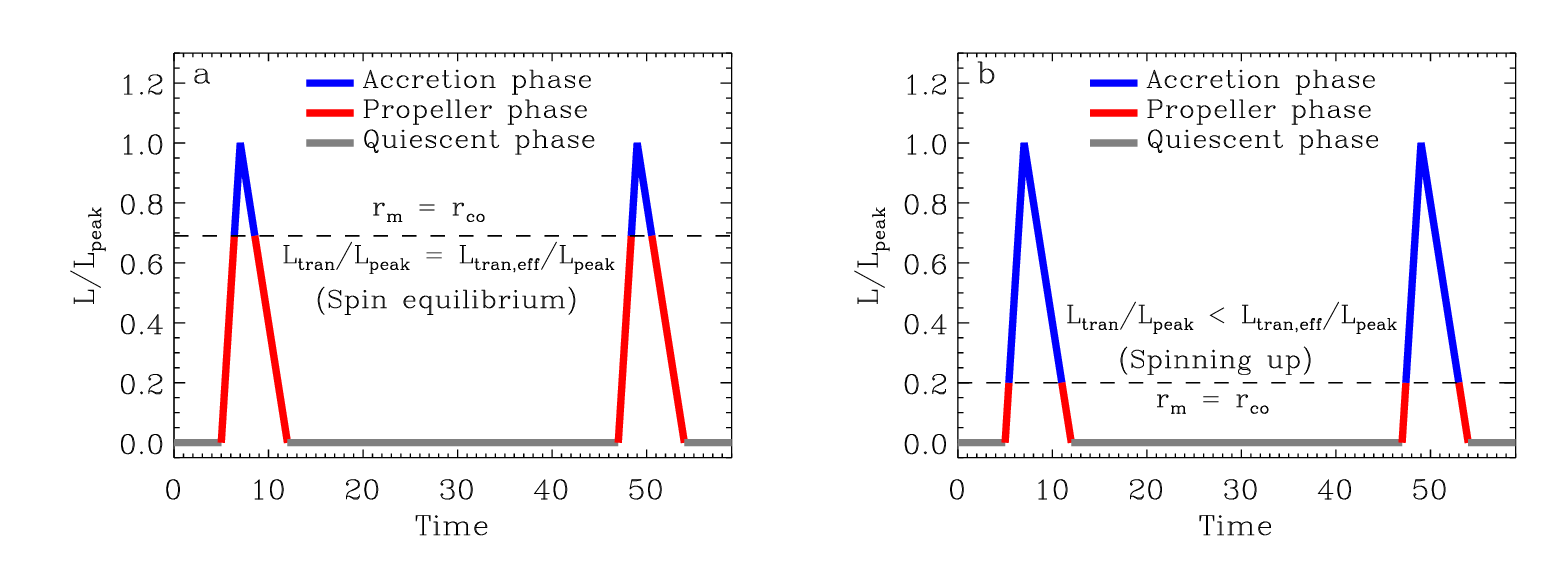}
\caption{Schematic illustration of three phases of an outburst cycle of a transient source.
	Here, we assume a linear luminosity ($L$) profile with the same peak value ($L_{\rm peak}$)
for each outburst. Time is given in an arbitrary unit and two outbursts are shown in each 
panel. {\it Panel a} is for the effective spin equilibrium of the neutron star, 
as the total positive angular momentum transfer to the star in the accretion phase (blue 
portion) balances the total negative angular momentum transfer in the propeller phase (red
portion). This happens if the luminosity $L_{\rm tran}$ corresponding to the transition
between accretion and propeller phases (defined by $r_{\rm m} = r_{\rm co}$) is equal to
$L_{\rm tran,eff} = 0.69\times L_{\rm peak}$ (see Section~\ref{previous}). 
{\it Panel b} is similar to the {\it panel a}, but here the star is still spinning up and 
has not yet reached the effective spin equilibrium. This happens when $L_{\rm tran}$ is
less than $L_{\rm tran,eff}$ (see Section~\ref{samepeak}). 
\label{fig1}}
\end{figure}

Now we discuss spin equilibrium of neutron stars in transient systems by disk--magnetosphere interaction alone.
Equation~\ref{equilibrium} does not give the spin equilibrium frequency
for a transient accretor.
This is because the $r_{\rm m} = r_{\rm co}$ condition cannot be satisfied
throughout an outburst, as $\dot{M}$, and hence $r_{\rm m}$, drastically evolve.
A practical way to define the spin equilibrium for a transient is by considering that no net angular
momentum is transferred to the neutron star in an outburst cycle. \citet{BhattacharyyaChakrabarty2017}
have shown that this criterion for spin equilibrium works well for transients, as the corresponding
effective spin equilibrium frequency matches within a few percent with the numerical result.
As shown in Equation~20 of \citet{BhattacharyyaChakrabarty2017}, the criterion of `no net angular
momentum transfer' implies 
\begin{equation}\label{balance1}
	L_{\rm peak}^{13/7} - L_{\rm tran,eff}^{13/7} = L_{\rm tran,eff}^{13/7}.
\end{equation}
Here, and in Figure~\ref{fig1}a which explains this equation, $L_{\rm peak}$ is the peak luminosity 
of each outburst with a linear $L$ profile, $L_{\rm tran}$ is 
the luminosity corresponding to the transition between accretion and propeller phases
(i.e., $r_{\rm m} = r_{\rm co}$) at the present time, and $L_{\rm tran,eff}$ is
the $L_{\rm tran}$ value when the effective spin equilibrium is reached.
The left-hand side of Equation~\ref{balance1} is proportional to the 
stellar angular momentum gain in the accretion phase (blue portion of an outburst in Figure~\ref{fig1}a),
while the right-hand side of the same equation 
is proportional to the stellar angular momentum loss in the propeller phase (red portion of an outburst 
in Figure~\ref{fig1}a). Equation~\ref{balance1} implies
\begin{equation}\label{Eff1} 
	l_{\rm eff} = \frac{L_{\rm tran,eff}}{L_{\rm peak}} = 2^{-7/13} = 0.69,
\end{equation}
in the effective spin equilibrium, which is shown by a dashed horizontal line in Figure~\ref{fig1}a.

Note that the luminosity $L$ and hence $r_{\rm m}$ evolve for a transient source, resulting in
an equilibrium spin frequency $\nu_{\rm eq}$ value corresponding to each luminosity value during 
an outburst cycle (see Equation~\ref{equilibrium} and $L \propto \dot{M}$). 
Two such frequencies are $\nu_{\rm eq,tran,eff}$ and $\nu_{\rm eq,peak}$, which are
	equilibrium spin frequencies that would be obtained (using Equation~\ref{equilibrium}) in case of persistent accretion
corresponding to $L_{\rm tran,eff}$ and $L_{\rm peak}$ respectively.
Therefore, using Equations~\ref{Eff1} and \ref{equilibrium} we get
\begin{equation}\label{Eff2} 
	\frac{\nu_{\rm eq,tran,eff}}{\nu_{\rm eq,peak}} = 
	\left[\frac{L_{\rm tran,eff}}{L_{\rm peak}}\right]^{3/7} = 2^{-3/13} = 0.85,
\end{equation}
We note that $\nu_{\rm eq,tran,eff}$, being the equilibrium spin frequency (Equation~\ref{equilibrium}) corresponding to 
	$L_{\rm tran,eff}$, is also the effective spin equilibrium frequency for a transient source, 
as defined in \citet{BhattacharyyaChakrabarty2017}.
Therefore, with mass transfer, $\nu$ increases and tends to become $\nu_{\rm eq,tran,eff}$ for a transient source,
as this effective spin equilibrium frequency is the maximum spin frequency a transiently accreting neutron
star can attain by disk--magnetosphere interaction alone \citep{BhattacharyyaChakrabarty2017}.

We sometimes consider two additional spin-down torques in this paper.
If the neutron star loses angular momentum because of the electromagnetic torque 
\begin{equation}\label{EMTorque}
	N_{\rm EM} = - \frac{2\mu^2}{3r_{\rm lc}^3} = - \frac{16\pi^3\mu^2\nu^3}{3c^3}
\end{equation}
due to magnetic dipole radiation during the quiescence period, 
the effective spin equilibrium frequency will be smaller than $\nu_{\rm eq,tran,eff}$.
Here, the speed-of-light cylinder radius $r_{\rm lc} = c/2\pi\nu$.
Besides, if the neutron star loses angular momentum continuously
due to gravitational wave torque
\begin{equation}\label{GWTorque}
	N_{\rm GW} = - \frac{32GQ^2}{5}\left(\frac{2\pi\nu}{c}\right)^5,
\end{equation}
the effective spin equilibrium frequency will also be smaller than $\nu_{\rm eq,tran,eff}$.
Here, $Q$ is the stellar rotating misaligned mass quadrupole moment \citep{Bildsten1998}.

\section{A new way to test spin equilibrium in transients}\label{method}

\subsection{For outbursts with the same peak luminosity}\label{samepeak}

In this section, we use the background given in Sections~\ref{Introduction} and \ref{previous}
to describe a way to find out if a neutron star in a transient system
has reached the effective spin equilibrium by disk--magnetosphere interaction alone.
We assume linear luminosity profiles of outbursts with the same peak luminosity $L_{\rm peak}$ for each outburst.
For our purpose, we use only one measurable parameter, i.e., the ratio of two source
luminosities $l = L_{\rm tran}/L_{\rm peak}$ (see Section~\ref{previous}).
When the effective spin equilibrium is reached, i.e., $L_{\rm tran} = L_{\rm tran,eff}$,
then $l = l_{\rm eff} = L_{\rm tran,eff}/L_{\rm peak} = 0.69$ (Section~\ref{previous}).
Therefore, if the measured value of $l$ is consistent with $l_{\rm eff}$ ($= 0.69$), one can conclude that
the neutron star  has reached the effective spin equilibrium by disk--magnetosphere interaction.
Note that $l_{\rm eff}$ is the maximum value that $l$ can achieve, because the effective spin equilibrium
frequency $\nu_{\rm eq,tran,eff}$ is the maximum spin frequency a neutron star can attain
while spinning up via disk--magnetosphere interaction (Section~\ref{previous}).
Therefore, a lower value of $l$ implies that a net positive angular momentum is being 
transferred to the neutron star (see Figure~\ref{fig1}b) and the star is still spinning up.
Hence, if the measured $l$ value is significantly less than $l_{\rm eff}$ ($= 0.69$),
one can conclude that the neutron star has not yet reached the effective
spin equilibrium by disk--magnetosphere interaction.

Now, suppose the measured $l$ value indicates that the effective spin equilibrium has not been reached yet.
How can one then estimate the effective spin equilibrium frequency $\nu_{\rm eq,tran,eff}$,
if the stellar spin frequency $\nu$ is known? Note that, since
$L_{\rm tran}$ is the luminosity corresponding to the transition between accretion 
and propeller phases (i.e., $r_{\rm m} = r_{\rm co}$) at the present time and $\nu$ is the current stellar spin frequency
(see Sections~\ref{Introduction} and \ref {previous}), the equilibrium spin frequency $\nu_{\rm eq}$ 
(Equation~\ref{equilibrium}) corresponding to $L_{\rm tran}$ is $\nu$. Therefore,
\begin{equation}\label{spin1}
	\nu \propto B^{-6/7} R^{-18/7} M^{5/7} L_{\rm tran}^{3/7}.
\end{equation}
On the other hand (see Section~\ref{previous}),
\begin{equation}\label{spin2}
	\nu_{\rm eq,tran,eff} \propto B_{\rm eff}^{-6/7} R_{\rm eff}^{-18/7} M_{\rm eff}^{5/7} L_{\rm tran,eff}^{3/7},
\end{equation}
where, $B_{\rm eff}$, $R_{\rm eff}$ and $M_{\rm eff}$ are stellar magnetic field, radius, and mass 
when the effective spin equilibrium is reached. Therefore,
\begin{equation}\label{spin3}
	\nu_{\rm eq,tran,eff} = \nu \left(\frac{B}{B_{\rm eff}}\right)^{6/7} \left(\frac{R}{R_{\rm eff}}\right)^{18/7} \left(\frac{M_{\rm eff}}{M}\right)^{5/7} \left(\frac{l_{\rm eff}}{l}\right)^{3/7}.
\end{equation}
Note that, since the neutron star will reach the spin equilibrium via accretion-induced spin-up, $M_{\rm eff} > M$.
Considering a neutron star mass range of $(1-2) M_\odot$, and a mass increase of $(0.1-0.4) M_\odot$
due to accretion, the range of $M_{\rm eff}/M$ is $1.05-1.4$. The fractional change of $R$ is usually much smaller than
the fractional change of $M$. Therefore, it is reasonable to consider $R^{18/7}{M_{\rm eff}}^{5/7} > 
{R_{\rm eff}}^{18/7}M^{5/7}$. We also note that, for fast spinning neutron stars,
it is reasonable to assume a fixed magnetic field strength, which is already low \citep{BhattacharyyaChakrabarty2017}.
This implies $B = B_{\rm eff}$. However, even if there is a reduction of the magnetic field value due to accretion, then
$B > B_{\rm eff}$. Therefore, 
\begin{equation}\label{spin4}
	\nu_{\rm eq,tran,eff} \ge \nu (l_{\rm eff}/l)^{3/7}.
\end{equation}
Thus, the ratio of two measured luminosities ($L_{\rm peak}$ and $L_{\rm tran}$)
can be used to test if a neutron
star has reached the effective spin equilibrium, and additionally the measured stellar spin frequency can 
provide a lower limit of the effective spin equilibrium frequency.

\subsection{For outbursts with varying peak luminosity}\label{diffpeak}

In Section~\ref{samepeak}, we assumed the same peak luminosity $L_{\rm peak}$ for each outburst.
But, in reality, $L_{\rm peak}$ can drastically vary from one outburst to another. Let us now explore
how to incorporate a varying $L_{\rm peak}$ in our method.

Suppose, $L_{\rm peak}^{\rm max}$ and $L_{\rm peak}^{\rm min}$ are maximum and minimum values of $L_{\rm peak}$ 
respectively, and $L_{\rm peak}$ varies in this range.
In this scenario, the effective spin equilibrium is reached if no net angular momentum is transferred
to the neutron star during a set of large numbers of outbursts. Note that it is quite practical to define an
effective spin equilibrium in this way, because given the small duraton (e.g., months) of each outburst cycle, the
total duration of a large number of outbursts is much smaller compared to the spin-up time scale (typically
$> 10^8$ years; see Figures~\ref{fig2} and \ref{fig3}). 

Here we write the angular momentum balance equation for a varying peak luminosity after
generalizing Equation~\ref{balance1} by summing over a large number ($K$) of outbursts:
\begin{equation}\label{balance2}
	\sum_{s>l_{\rm m,eff}} \left[s^{13/7} - l_{\rm m,eff}^{13/7}\right] = \sum_{s>l_{\rm m,eff}} l_{\rm m,eff}^{13/7} + \sum_{s \le l_{\rm m,eff}} s^{13/7}.
\end{equation}
Here, $s = L_{\rm peak}/L_{\rm peak}^{\rm max}$ and $l_{\rm m,eff} = L_{\rm tran,eff}/L_{\rm peak}^{\rm max}$.
Note that the left-hand side of Equation~\ref{balance2} is proportinal to the stellar angular
momentum gain for the outbursts ($k$ in number) with $s>l_{\rm m,eff}$ (or, $L_{\rm peak} > L_{\rm tran,eff}$), 
for which the accretion phase exists when spin equilibrium is reached.
The right-hand side of Equation~\ref{balance2} is proportinal to the stellar angular momentum loss in the propeller phase.
Note that the second term on the right-hand side is for $(K-k)$ number of outbursts with
$s \le l_{\rm m,eff}$ ($L_{\rm peak} \le L_{\rm tran,eff}$), for which only the propeller phase exists. 
Note that Equation~\ref{balance2} is valid for any distribution of $L_{\rm peak}$ with 
$L_{\rm peak} \le L_{\rm peak}^{\rm max}$. Equation~\ref{balance2} gives
\begin{equation}\label{balance3}
	l_{\rm m,eff} = \left[\frac{1}{2k}\left(\sum_{s>l_{\rm m,eff}} s^{13/7} - \sum_{s \le l_{\rm m,eff}} s^{13/7}\right)\right]^{7/13}.
\end{equation}
Thus we can estimate $l_{\rm m,eff}$, which is a generalized form of $l_{\rm eff}$ (Equation~\ref{Eff1})
for the following reason. $l_{\rm eff}$ is defined for the same $L_{\rm peak}$ value for every outburst 
(Section~\ref{previous}), which implies $L_{\rm peak}^{\rm max} = L_{\rm peak}$ and $L_{\rm peak} > L_{\rm tran,eff}$. 
The former gives $s = 1$, and hence $\sum_{s>l_{\rm m,eff}} s^{13/7} = k$, while the latter gives 
$K-k = 0$, and hence $\sum_{s \le l_{\rm m,eff}} s^{13/7} = 0$.
This means $l_{\rm m,eff}$ in Equation~\ref{balance3} reduces to $2^{-7/13}$, i.e., 
the expression of $l_{\rm eff}$ given in Equation~\ref{Eff1}.

Therefore, as discussed in Section~\ref{samepeak}, if the measured $l_{\rm m} (= L_{\rm tran}/L_{\rm peak}^{\rm max})$ 
is significantly less than $l_{\rm m,eff}$, then the neutron star has not yet reached the effective
spin equilibrium by disk--magnetosphere interaction alone.
In this case, we can write 
\begin{equation}\label{new2}
	\nu_{\rm eq,tran,eff} \ge \nu \left(l_{\rm m,eff}/l_{\rm m}\right)^{3/7},
\end{equation}
by suitably modifying Equation~\ref{spin4}.

\subsection{Application to Aql X--1}\label{Application}

Now we apply the above method to Aql X--1. This is an ideal transient neutron star LMXB for this 
purpose, because it shows frequent outbursts, and its $L_{\rm tran}$ value has been reported.
We consider $L_{\rm peak}^{\rm max} \approx 8\times10^{37}$~erg s$^{-1}$ \citep{Kitamotoetal1993}
and $L_{\rm peak}^{\rm min} \approx 2\times10^{36}$~erg s$^{-1}$ \citep{Campanaetal2013}.
We choose the $L_{\rm peak}$ value randomly from this wide range for each outburst.
Two different values of $L_{\rm tran}$ have been reported using two methods: $1.3\times10^{36}$~erg s$^{-1}$ \citep{Asaietal2013}
and $(5.3-7.5)\times10^{36}$~erg s$^{-1}$ \citep[after accounting for a source distance of 5 kpc; ][]{Campanaetal2014}.
Note that, while both of these methods rely on the expectation that an accretion phase to propeller phase
transition causes a quick X-ray luminosity fall, the former method identifies a two-step fall based
on the spectral analysis, considers that only the second step is due to the accretion-to-propeller 
transition, and thus infers a lower $L_{\rm tran}$ value.
We add an ad hoc 20\% uncertainty to the first $L_{\rm tran}$ value ($1.3\times10^{36}$~erg s$^{-1}$) 
to be conservative, and thus consider a range 
of $(1.04-1.56)\times10^{36}$~erg s$^{-1}$. These give the $l_{\rm m}$ $(= L_{\rm tran}/L_{\rm peak}^{\rm max})$ 
ranges of $0.013-0.0195$ and $0.066-0.094$ corresponding to the $L_{\rm tran}$ values reported by
\citet{Asaietal2013} and \citet{Campanaetal2014} respectively.
On the other hand, solving Equation~\ref{balance3} by numerical iterations for the above-mentioned
$L_{\rm peak}^{\rm max}$ and $L_{\rm peak}^{\rm min}$ values of Aql X--1, we find 
$l_{\rm m,eff} = 0.478$. Therefore, $l_{\rm m}$ is significantly less than $l_{\rm m,eff}$, and hence 
we can conclude that the neutron star
in Aql X--1 has not yet reached the effective spin equilibrium by disk--magnetosphere interaction alone.
Besides, since $\nu = 550$~Hz for Aql X--1 \citep{PatrunoWatts2012}, the lower limits of the effective spin 
equilibrium frequency $\nu_{\rm eq,tran,eff}$ is in the ranges $2167-2578$~Hz and $1104-1285$~Hz for
the measurements of \citet{Asaietal2013} and \citet{Campanaetal2014} respectively. This means,
in the absence of an additional spin-down mechanism (apart from the propeller effect)
and after sufficient mass transfer, the neutron star in 
Aql X--1 not only can become a submillisecond pulsar, but also may reach the breakup spin rate limit
\citep[e.g., ][]{Bhattacharyyaetal2016}.
However, since no submillisecond pulsar has been detected so far, the existence of 
one or more additional spin-down mechanisms is plausible. Note that nothing is known on the 
long-term spin evolution of Aql X--1. If in the future it is found that the neutron star 
in Aql X--1 is not overall spinning up, then that would be an evidence of one or more 
additional spin-down mechanisms (e.g., electromagnetic radiation, gravitational waves).

The  $L_{\rm peak}$ distribution for Aql X--1 is not known, and hence it is reasonable to 
assume a random distribution between $L_{\rm peak}^{\rm min}$ and $L_{\rm peak}^{\rm max}$.
If, in reality, $L_{\rm peak}$ systematically has somewhat higher
values (for a fixed $L_{\rm peak}^{\rm max}$ value), the $l_{\rm m,eff}$ value will usually be higher,
resulting in an even larger effective spin equilibrium frequency for Aql X--1. On the other hand, if 
$L_{\rm peak}$ systematically has somewhat lower values, the effective spin equilibrium frequency for Aql X--1
could be lower. However, our conclusion, that Aql X--1 has not yet reached the effective spin equilibrium
by disk--magnetosphere interaction alone, should be reliable, because a clear systematic behavior of 
long-term $L_{\rm peak}$ distribution is not known, and we consider a large range of $L_{\rm peak}$.

What could be the implication of our assumption of linear luminosity profiles of outbursts? Note that 
Aql X--1 can indeed have fairly linear outburst profiles \citep[e.g., see Figure 2 of ][]{Shahbazetal1998}.
However, some outburst profiles of the source show a tendency of flatness near the peak \citep[e.g., ][]{Gungoretal2014}.
But this implies an even larger effective spin equilibrium frequency, as the source spends
more time in the accretion phase, when a positive angular momentum is transferred to the neutron star 
\citep{BhattacharyyaChakrabarty2017}. Therefore, it can be concluded that the neutron star in Aql X--1 has not yet
reached the effective spin equilibrium by disk--magnetosphere interaction alone considering realistic outburst light curves. 

\section{Numerical computation for Aql X--1 without gravitational wave torque}\label{noGW}

In Section~\ref{Application}, from a semi-analytical angular momentum balance study, we showed that
the neutron star in Aql X--1 is still spinning up toward the effective spin equilibrium
if only disk--magnetosphere interaction is responsible for spin evolution.
This means if this neutron star has already reached an effective
spin equilibrium in reality (although it is not known), then at least one additional spin-down mechanism is at play.
In this section, we perform detailed numerical computations of the spin evolution of this source through 
a series of outbursts, using the disk--magnetosphere interaction torques given in Equations~\ref{Torque7} and \ref{Torque8},
and confirm the conclusion of Section~\ref{Application}. Next, we repeat these numerical computations
with an additional spin-down due to the electromagnetic torque (Equation~\ref{EMTorque}) during the 
quiescence periods (but not including the gravitational wave torque). 
The procedure of these numerical computations is the same as described in 
\citet{BhattacharyyaChakrabarty2017}, except here we choose the peak accretion rate ($\dot{M}_{\rm peak}$) of an outburst 
randomly from a range between maximum ($\dot{M}_{\rm peak}^{\rm max}$) and minimum ($\dot{M}_{\rm peak}^{\rm min}$)
values, as mentioned in Section~\ref{Application}. 
Note that one needs to use accretion rate instead of luminosity for numerical computations of neutron star spin evolution.

For numerical computations, we use $\dot{M}_{\rm peak}^{\rm min}/\dot{M}_{\rm peak}^{\rm max} = 0.025$ (as observed 
$L_{\rm peak}^{\rm min}/L_{\rm peak}^{\rm max} = 0.025$ for Aql X--1; Section~\ref{Application}), and use three
values of the long-term average accretion rate $\dot M_{\rm av}$: $5\times10^{15}$~g s$^{-1}$, 
$2.65\times10^{16}$~g s$^{-1}$ and $5\times10^{16}$~g s$^{-1}$. This wide range is consistent with an
estimated $\dot M_{\rm av}$ value of $7\times10^{15}$~g s$^{-1}$ for Aql X--1 \citep{Campanaetal2013}. We use 
three values of average $\dot{M}_{\rm peak}/\dot M_{\rm av}$ for each $\dot M_{\rm av}$ value, so that
the corresponding outburst duty cycle is consistent with that observed \citep[see ][]{Campanaetal2013}.
Besides, we use three values of $\xi$ (0.5, 1.0, 1.4), which are in the range ($0.5-1.4$) suggested by many 
previous works, some of which used simulations \citep[e.g., ][]{GhoshLamb1979, Wang1996, Longetal2005}. 
We also note that a $\xi$ value outside this range \citep[as indicated in][]{Ertan2017} 
does not affect our overall results (as long as 
there is an interaction between the disk and the magnetosphere), but can only affect inferred constraints on other
parameters \citep[for example, on stellar magnetic field as $B \propto \xi^{-7/4}$; ][]{BhattacharyyaChakrabarty2017}.
We use two widely different values of $\eta$ \citep[0.2, 1; ][]{BhattacharyyaChakrabarty2017} to make our results
robust, and three values of the stellar initial mass ($1.1 M_\odot$, $1.35 M_\odot$, $1.6 M_\odot$) 
in a reasonable and large range. Since we aim to constrain the stellar magnetic field, we use a large number 
(50) of values of $B$ in the range of $1\times10^7$~G $-$ $5\times10^8$~G
\citep{Asaietal2013, Campanaetal2014}. Therefore, we have $3\times3\times3\times2\times3\times50 = 8100$
parameter combinations, and we numerically compute the spin evolution for each combination.

\begin{figure}[h]
\vspace{-1.0cm}
\hspace{-1.0cm}
\includegraphics*[width=18.0cm]{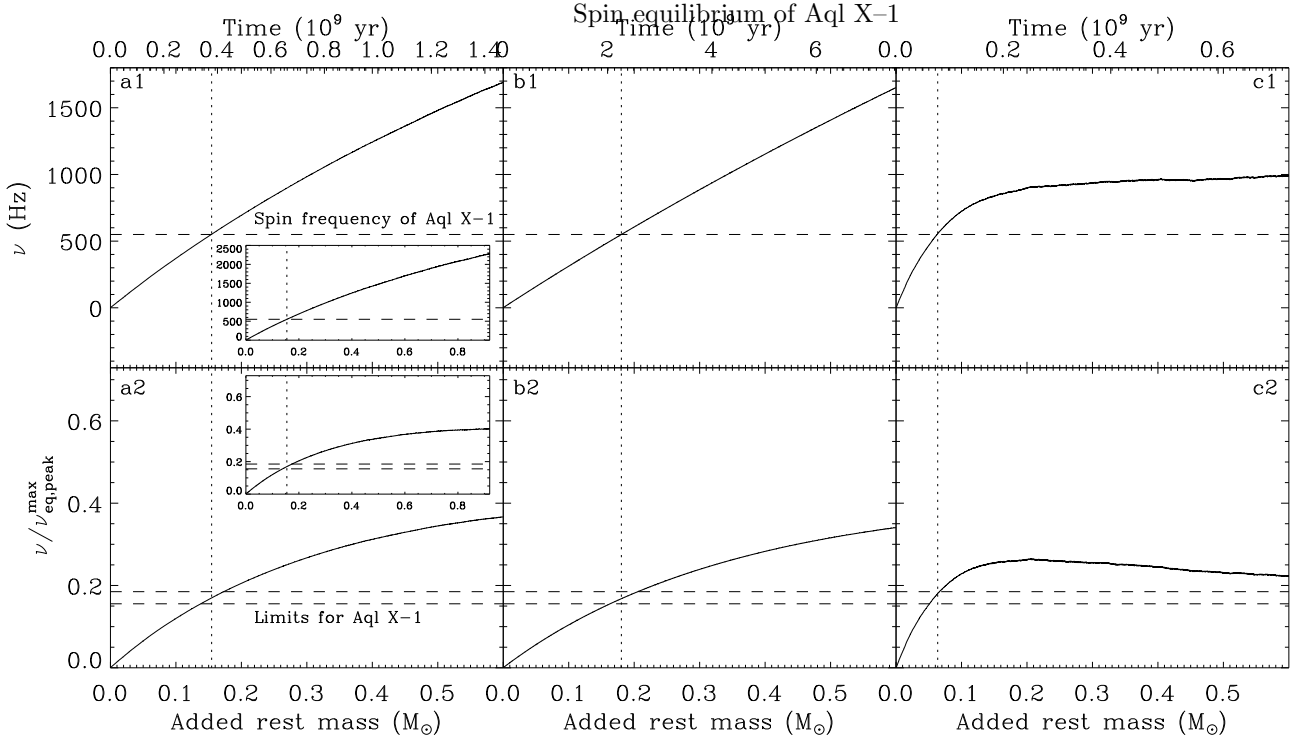}
\caption{Three examples of neutron star spin evolution curves for Aql X--1, which are allowed by observations,
	considering spin-down due to electromagnetic radiation, but without spin down due to gravitational waves
	(see Section~\ref{noGW}). The upper panel curves show the spin frequency ($\nu$ in Hz) versus the 
	rest mass transferred to neutron star (upper x-axes show time).
	The lower panels show the same curves, but with $\nu$ in the unit of the spin equilibrium frequency 
	$\nu_{\rm eq,peak}^{\rm max}$ corresponding to the maximum value $\dot{M}_{\rm peak}^{\rm max}$ of 
	the outburst peak accretion rates.
	{\it Panels a1/a2}, {\it panels b1/b2} and {\it panels c1/c2} are for three different sets of values of parameters
	($\xi$, $\eta$, $B$, initial $M$, $\dot{M}_{\rm av}$, average $\dot{M}_{\rm peak}/\dot{M}_{\rm av}$), viz.,
	(1.0, 0.2, $4.0\times10^7$~G, 1.35~$M_\odot$, $2.65\times10^{16}$~g s$^{-1}$, 10), 
	(1.4, 0.2, $1.0\times10^7$~G, 1.1~$M_\odot$, $5.0\times10^{15}$~g s$^{-1}$, 10), 
	and (0.5, 0.2, $3.4\times10^8$~G, 1.6~$M_\odot$, $5.0\times10^{16}$~g s$^{-1}$, 20) respectively.
	The dashed horizontal line in the upper panels indicates the current $\nu$-value (550 Hz) of Aql X--1,
	while the pair of dashed horizontal lines in the lower panels give an observationally inferred range
	($0.155-0.185$) of $\nu/\nu_{\rm eq,peak}^{\rm max}$. Each of the above three sets of parameter values is allowed
	for Aql X--1, because the spin evolution curve simultaneously satisfies both the current $\nu$-value and 
	the above-mentioned inferred $\nu/\nu_{\rm eq,peak}^{\rm max}$ range of Aql X--1, as shown by a
	dotted vertical line. Besides, {\it Panels a1/a2} have insets showing the same curves in the corresponding panels,
	but extended up to close to the spin equilibrium, which gives an idea of how much mass has to be accreted 
	to attain equilibrium for this set of parameter values.
\label{fig2}}
\end{figure}

\begin{figure}[h]
\centering
\includegraphics*[width=7.0cm]{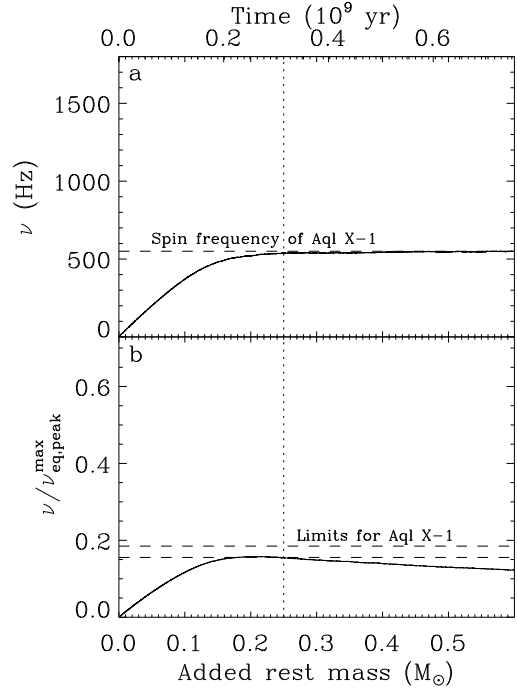}
\caption{Similar to Fig.~\ref{fig2}, but the numerical computation includes spin down due to gravitational 
	radiation (Section~\ref{GW}).
	The parameter values ($\xi$, $\eta$, $Q$, $B$, initial $M$, $\dot{M}_{\rm av}$, average
	$\dot{M}_{\rm peak}/\dot{M}_{\rm av}$) are (1.0, 0.2, $1.0\times10^{37}$~g cm$^2$, $4.0\times10^7$~G, 1.1~$M_\odot$, $5.0\times10^{16}$~g s$^{-1}$, 5).
\label{fig3}}
\end{figure}

Note that the two observationally inferred ranges of $l_{\rm m}$ $(= L_{\rm tran}/L_{\rm peak}^{\rm max})$, i.e., 
$0.013-0.0195$ and $0.066-0.094$ (see Section~\ref{Application}) imply $0.155-0.185$
and $0.312-0.363$ ranges of $\nu/\nu_{\rm eq,peak}^{\rm max}$ ($= l_{\rm m}^{3/7}$; Equation~\ref{equilibrium}) 
respectively. This is because $\nu \propto L_{\rm tran}^{3/7}$ (Equation~\ref{spin1}), and the 
equilibrium spin frequency $\nu_{\rm eq,peak}^{\rm max}$ corresponding to the maximum peak luminosity 
$L_{\rm peak}^{\rm max}$ is proportonal to ${L_{\rm peak}^{\rm max}}^{3/7}$ (using Equation~\ref{equilibrium}). 
Now, how do we determine which of the above-mentioned 8100 parameter combinations are allowed for Aql X--1?
For this, we numerically compute the spin evolution for each combination until $0.6 M_\odot$ rest mass is
transferred to the neutron star. If, during such an evolution, two observed values, viz., $\nu = 550$~Hz
and the observationally inferred $\nu/\nu_{\rm eq,peak}^{\rm max}$ range, can be simultaneously obtained at any 
point in time, then the corresponding parameter combination is allowed for Aql X--1.
In Figure~\ref{fig2}, we give examples of spin evolution for three such allowed
parameter combinations (for $0.155 < \nu/\nu_{\rm eq,peak}^{\rm max} < 0.185$ at $\nu \approx 550$~Hz)
with widely different parameter values. 
By identifying all of the allowed parameter combinations from our 8100 combinations,
we find out if the neutron star in Aql X--1
has reached the effective spin equilibrium, and constrain the stellar magnetic field, as discussed below.

An indicator of the effective spin equilibrium is the nature of the $\nu$ evolution curve. 
The $\nu$ value increases rapidly before this equilibrium is reached, and after this $\nu$
(and the effective spin equilibrium frequency) evolves slowly as the neutron star mass increases by accretion
\citep[e.g., Figure~2 of ][]{BhattacharyyaChakrabarty2017}.
But a clearer indicator of the effective spin equilibrium is the evolution curve of $\nu$ in the unit of the
equilibrium spin frequency corresponding to the outburst peak luminosity. If we consider spin evolution
by disk--magnetosphere interaction alone, the above-mentioned curve saturates when effective spin equilibrium
is reached, as can be seen from Figure~6b of \citet{BhattacharyyaChakrabarty2017}. But if we include
an additional spin-down mechanism (for example, due to electromagnetic torque), the above-mentioned curve
attains a maximum roughly when effective spin equilibrium is reached, and then decreases, as can be seen from the 
{\it panel c2} of Figure~\ref{fig2}.
This curve keeps on increasing before the effective spin equilibrium is reached, as can be seen 
from the {\it panels a2} and {\it b2} of Figure~\ref{fig2}.
Note that, while we compute spin evolution till $0.6 M_\odot$ rest mass is transferred for all cases,
in one case (insets of {\it panels a1} and {\it a2} of Figure~\ref{fig2}), we compute up to
a large ($> 2000$~Hz) $\nu$-value to give an idea of how much mass has to be typically accreted to attain the
spin equilibrium, when the equilibrium spin frequency is very high.

Using the above indicators, we find from our numerical computations that the neutron star of Aql X--1 has not yet reached
the effective spin equilibrium for any parameter combination, if we consider spin evolution
by disk--magnetosphere interaction alone. This is consistent with the semi-analytical result reported in 
Section~\ref{Application} and validates the new method described in Section~\ref{method}. 
Even when we consider an additional spin-down due to electromagnetic torque and 
an observationally inferred range $0.155-0.185$ for $\nu/\nu_{\rm eq,peak}^{\rm max}$, no parameter combination is found
for which the effective spin equilibrium is reached. This indicates if $0.155-0.185$ is the correct
range of $\nu/\nu_{\rm eq,peak}^{\rm max}$, the neutron star is still spinning up. 
But if we consider the electromagnetic torque and an observationally inferred range $0.312-0.363$
for $\nu/\nu_{\rm eq,peak}^{\rm max}$, the effective spin equilibrium is reached
for a small fraction of parameter combinations. Therefore, even with the electromagnetic spin-down 
torque, while it cannot be ruled out that the effective spin equilibrium has been reached,
it is more likely that the neutron star in Aql X--1 is still spinning up.
This can be tested if the long-term spin evolution of Aql X--1 is measured in the future.

Next, we constrain the stellar magnetic field $B$ using the allowed parameter combinations for Aql X--1.
We find that, considering the additional spin-down due to the electromagnetic torque,
the stellar magnetic field $B$ can be constrained to the
ranges $1\times10^7$~G $-$ $3.4\times10^8$~G and $2\times10^7$~G $-$ $5\times10^8$~G
for  $\nu/\nu_{\rm eq,peak}^{\rm max} = 0.155-0.185$ and $0.312-0.363$ (observationally 
inferred ranges) respectively.
Note that $B$ strongly depends on $\xi$ \citep[$B \propto \xi^{-7/4}$; ][]{BhattacharyyaChakrabarty2017}, 
and hence a more constrained $\xi$ value 
from a better understanding of disk--magnetosphere interaction will be useful to constrain $B$
much more tightly. Knowledge of other parameters, such as the initial $M$ value, can also provide
significantly tighter constraints. For example, for $\xi = 1.0$ and the initial $M = 1.35 M_\odot$
\citep{ThorsettChakrabarty1999}, the above-mentioned constraints on $B$ reduce to
$3\times10^7$~G $-$ $8\times10^7$~G and $4\times10^7$~G $-$ $1.8\times10^8$~G respectively.

\section{Numerical computation for Aql X--1 with gravitational wave torque}\label{GW}

In order to check if Aql X--1 can emit gravitational radiation, we consider an additional
spin-down due to the gravitational wave torque (Equation~\ref{GWTorque}). For this, we use
45 non-zero values of the neutron star misaligned mass quadrupole moment $Q$ up to $10^{38}$ g cm$^2$.
Therefore, we use $8100\times45 = 36,4500$ additional parameter combinations to compute the spin evolution
(see Section~\ref{noGW}). In the same way as described in Section~\ref{noGW}, we can find 
which of these parameter combinations are allowed for Aql X--1. This can be useful to constrain parameter values.
Figure~\ref{fig3} depicts the spin evolution curve for one such allowed parameter combination,
which shows that Aql X--1 may emit gravitational radiation. This figure also shows (see Section~\ref{noGW}) that 
the neutron star in Aql X--1 could be in the effective spin equilibrium, if it emits gravitational radiation.

We find that, for our parameter combinations (as mentioned in Section~\ref{noGW}), the maximum allowed $Q$-value is 
$1.6\times10^{37}$ g cm$^2$ for Aql X--1, which implies that this source may emit gravitational radiation. 
However, since the lower limit of $Q$ is zero, the gravitational radiation from Aql X--1 cannot be established 
with certainty. 

\begin{figure}[h]
\centering
\includegraphics*[width=10.0cm]{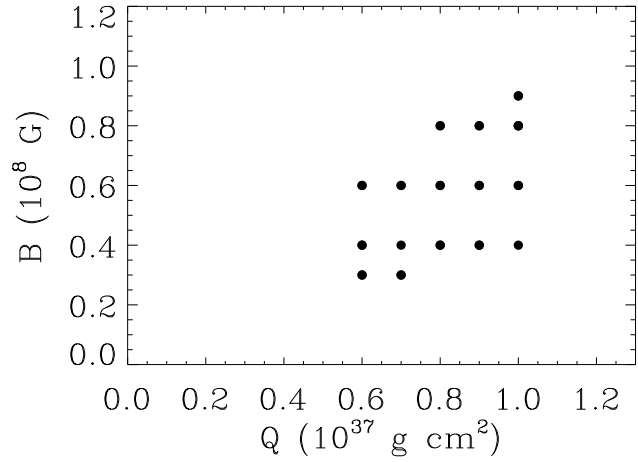}
\caption{Example of constraint on the $B$ versus $Q$ space: $B$ versus $Q$ 
points, which simultaneously satisfy a $\nu/\nu_{\rm eq,peak}^{\rm max}$ range of 
$0.155-0.185$ and $\nu = 550$~Hz (from numerical computation of spin evolution, including spin down due to gravitational radiation). Here, we use 50 $B$ values in the range of $(0.1-5.0)\times10^8$~G and 46 $Q$ values in the range of $(0.0-1.0)\times10^{38}$~g cm$^2$. We also use ranges of other parameters as mentioned in Section~\ref{noGW}, except here we consider $\xi = 1.0$ and $1.5 M_\odot \le M \le 1.6 M_\odot$ (with $M$ as the current neutron star mass at $\nu = 550$~Hz). Note that this assumed mass range 
is arbitrary, and not for Aql X--1. This figure demonstrates how $B$ and $Q$ could be tightly constrained using 
the knowledge of other parameter values (Section~\ref{GW}).
\label{fig4}}
\end{figure}

It is generally possible to constrain the $B-Q$ 
space for a source using computations similar to those reported in this paper. 
We demonstrate this with an example assuming further constraints on the measured stellar mass (not applicable
for Aql X--1) and the $\xi$-value 
(Figure~\ref{fig4}). Note that there are gaps in the constrained region in this figure  because of very few used
values of most of the parameters.
Figure~\ref{fig4} shows that, if other parameter values are known with sufficient accuracy,
$Q$ may have a non-zero lower limit, which implies gravitational radiation.
Alternatively, a possible future detection of gravitational radiation can be useful to 
constrain various source parameters, including $B$.

\section{Conclusion}\label{conclusion}

In this paper, we provide a new practical way, based on specific measurable luminosities, to find out if 
a neutron star in a transient LMXB has reached the spin equilibrium by disk--magnetosphere interaction alone, and if not,
to estimate this spin equilibrium frequency using the known stellar spin rate. 
This will be very useful to understand the spin distribution 
of MSPs, as well as the observed cutoff of their spin rates.

The method involves the measurement of $L_{\rm tran}$, the luminosity corresponding to the transition between 
accretion and propeller phases. Note that, like neutron star mass and spin rate, $L_{\rm tran}$ should not 
perceivably change from one outburst to another, and hence can be estimated from one outburst, and may be 
confirmed from other outbursts.

Applying our method to the transient LMXB Aql X--1, we show that its neutron star has not yet 
reached the spin equilibrium by disk--magnetosphere interaction alone, and this spin equilibrium frequency
is more than a thousand Hz. While we cannot be definite about the gravitational wave emission from 
Aql X--1 from numerical computations, our numerical results
compared with the known stellar spin frequency and an observed luminosity ratio show that gravitational radiation 
from Aql X--1 is possible, with a $1.6\times10^{37}$ g cm$^2$ upper limit of the 
stellar misaligned mass quadrupole moment $Q$.
Note that this is not inconsistent with the inferred upper limit of $Q$ ($\le 2\times10^{36}$ g cm$^2$;
\citet{Papittoetal2011}; see also \citet{Patruno2010, Hartmanetal2011}) for a similarly
fast spinning MSP IGR J00291+5934 ($\nu \approx 599$~Hz).
However, there can be practical obstacles, for example, related to regular monitoring of 
parameter evolution, to detect such gravitational radiation \citep[e.g., ][]{Wattsetal2008}.
Finally, we emphasize that our new method provides an independent way to check if spin equilibrium has been 
reached, if additional spin-down mechanisms (e.g., gravitational wave torque) are required,
and to constrain the source parameters. A future estimation of the long-term spin evolution of Aql X--1 may 
provide a complementary method to achieve these goals for this source.

\clearpage

\clearpage

\clearpage

\clearpage


\begin{thebibliography}{}

\bibitem[Alpar et al. (1982)]{Alparetal1982} Alpar, M. A., Cheng, A. F., Ruderman, M. A., \& Shaham, J. 1982, Nature, 300, 728

\bibitem[Andersson et al. (1999)]{Anderssonetal1999} Andersson, N., Kokkotas, K. D., Stergioulas, N. 1999, ApJ, 516, 307

\bibitem[Archibald et al.(2009)]{Archibaldetal2009} Archibald, A. M., Stairs, I. H., Ransom, S. M., et al. 2009, Science, 324, 1411

\bibitem[Asai et al. (2013)]{Asaietal2013} Asai, K., Matsuoka, M., Mihara, T., Sugizaki, M., \& Serino, M. 2013, ApJ, 773, 117

\bibitem[Bassa et al. (2014)]{Bassaetal2014} Bassa, C. G., Patruno, A., Hessels, J. W. T., et al. 2014, MNRAS, 441, 1825

\bibitem[Bhattacharyya \& Chakrabarty (2017)]{BhattacharyyaChakrabarty2017} Bhattacharyya, S., \& Chakrabarty, D. 2017, ApJ, 835, 4

\bibitem[Bhattacharyya et al.(2016)]{Bhattacharyyaetal2016} Bhattacharyya, S., Bombaci, I, Logoteta, D., Thampan, A. V. 2016, MNRAS, 457, 3101

\bibitem[Bildsten(1998)]{Bildsten1998} Bildsten, L. 1998, ApJL, 501, L89

\bibitem[Campana et al. (2014)]{Campanaetal2014} Campana, S., Brivio, F., Degenaar, N., Mereghetti, S., Wijnands, R., D'Avanzo, P., Israel, G. L., \& Stella, L. 2014, MNRAS, 441, 1984

\bibitem[Campana et al. (2013)]{Campanaetal2013} Campana, S., Coti Zelati, F., \& D'Avanzo, P. 2013, MNRAS, 432, 1695

\bibitem[Chakrabarty(2005)]{Chakrabarty2005} Chakrabarty, D. 2005, in Binary Radio Pulsars, ed. F. A. Rasio and I. H. Stairs, (ASP Conference Series: San Francisco), 328, 279

\bibitem[Chakrabarty \& Morgan(1998)]{ChakrabartyMorgan1998} Chakrabarty, D., \& Morgan, E. H. 1998, Nature, 394, 346

\bibitem[Chakrabarty et al.(2003)]{Chakrabartyetal2003} Chakrabarty, D., Morgan, E. H., Muno, M. P., et al. 2003, Nature, 424, 42

\bibitem[Cook et al. (1994)]{Cooketal1994} Cook, G. B., Shapiro, S. L., Teukolsky, S. A. 1994, ApJ, 424, 823

\bibitem[D'Angelo \& Spruit (2010)]{DAngeloetal2010} D'Angelo, C. R., \& Spruit, H. C. 2010, MNRAS, 406, 1208

\bibitem[de Martino et al.(2013)]{deMartinoetal2013} de Martino, D., Belloni, T., Falanga, M., et al. 2013, A\&A, 550, A89

\bibitem[Ertan (2017)]{Ertan2017} Ertan, \"Unal 2017, MNRAS, 466, 175

\bibitem[Ferrario \& Wickramasinghe(2007)]{FerrarioWickramasinghe2007} Ferrario, L., \& Wickramasinghe, D. 2007, MNRAS, 375, 1009

\bibitem[Ghosh \& Lamb(1979)]{GhoshLamb1979} Ghosh, P., \& Lamb, F. K. 1979, ApJ, 234, 296

\bibitem[G\"ung\"or et al. (2014)]{Gungoretal2014} G\"ung\"or, C., G\"uver, T., \& Eksi, Y. 2014, MNRAS, 439, 2717

\bibitem[Hartman et al. (2011)]{Hartmanetal2011} Hartman, J. M., Galloway, D. K., \& Chakrabarty, D. 2011, ApJ, 726, 26

\bibitem[Hessels(2008)]{Hessels2008} Hessels, J. W. T. 2008, AIP Conf. Proc., 1068, 130

\bibitem[Illarionov \& Sunyaev(1975)]{Illarionov1975} Illarionov, A.~F., \& Sunyaev, R.~A. 1975, A\&A, 39, 185

\bibitem[Kitamoto et al. (1993)]{Kitamotoetal1993} Kitamoto, S., Tsunemi, H., Miyamoto, S., \& Roussel-Dupre, D. 1993, ApJ, 403, 315

\bibitem[Lamb \& Yu(2005)]{LambYu2005} Lamb, F. K., \& Yu, W. 2005, in Binary Radio Pulsars, ed. F. A. Rasio \& I. H. Stairs, (ASP Conference Series: San Francisco), 328, 299

\bibitem[Long et al. (2005)]{Longetal2005} Long, M., Romanova, M. M., \& Lovelace, R. V. E. 2005, ApJ, 634, 1214

\bibitem[Papitto et al. (2013)]{Papittoetal2013} Papitto, A., Ferrigno, C., Bozzo, E., et al. 2013, Nature, 501, 517

\bibitem[Papitto et al. (2011)]{Papittoetal2011} Papitto, A., Riggio, A., Burderi, L., Di Salvo, T., D'Ai, A., \& Iaria, R. 2011, A\&A, 528, A55

\bibitem[Papitto et al.(2014)]{Papittoetal2014} Papitto, A., Torres, D. F., Rea, N., Tauris, T. M. 2014, A\&A, 566, A64

\bibitem[Patruno(2010)]{Patruno2010} Patruno, A. 2010, ApJ, 722, 909

\bibitem[Patruno et al.(2012)]{Patrunoetal2012b} Patruno, A., Haskell, B., \& D'Angelo, C. 2012, ApJ, 746, 9

\bibitem[Patruno \& Watts(2012)]{PatrunoWatts2012} Patruno, A., \& Watts, A. L. 2012, arXiv:1206.2727
	
\bibitem[Radhakrishnan \& Srinivasan(1982)]{RadhakrishnanSrinivasan1982} Radhakrishnan, V., \& Srinivasan, G. 1982, Current Science, 51, 1096

\bibitem[Rappaport et al.(2004)]{Rappaportetal2004} Rappaport, S. A., Fregeau, J. M., \& Spruit, H. 2004, ApJ, 606, 436

\bibitem[Shahbaz et al. (1998)]{Shahbazetal1998} Shahbaz, T., Bandyopadhyay, R. M., Charles, P. A., Wagner, R. M., Muhli, P., Hakala, P., Casares, J., \& Greenhill, J. 1998, MNRAS, 300, 1035

\bibitem[Thorsett \& Chakrabarty(1999)]{ThorsettChakrabarty1999} Thorsett, S. E., \& Chakrabarty, D. 1999, ApJ, 512, 288

\bibitem[Ustyugova et al.(2006)]{Ustyugova2006} Ustyugova, G.~V., Koldoba, A.~V., Romanova, M.~M., et al. 2006, ApJ, 646, 304

\bibitem[Wang (1996)]{Wang1996} Wang, Y.-M 1996, ApJL, 465, L111

\bibitem[Watts (2012)]{Watts2012} Watts, A. L. 2012, \araa, 50, 609

\bibitem[Watts et al. (2008)]{Wattsetal2008} Watts, A. L., Krishnan, B.,  Bildsten, L.,
Schutz, B. F. 2008, MNRAS, 389, 839

\bibitem[Wijnands \& van der Klis(1998)]{WijnandsKlis1998} Wijnands, R., \& van der Klis, M. 1998, Nature, 394, 344

\end{thebibliography}
\end{document}